%% file: main.tex
\pdfoutput=1
\documentclass[amsmath,amssymb,aps,prl,superscriptaddress,twocolumn,floatfix]{revtex4-2}
\usepackage{graphicx} 
\usepackage{wrapfig}
\usepackage{bm}
\usepackage[usenames]{color}
\usepackage{MnSymbol}
\usepackage{hyperref} 
\usepackage{subfigure}

\usepackage[dvipsnames]{xcolor}
\definecolor{darkblue}{HTML}{0b3873}

\newcommand{\para}{}

\usepackage[normalem]{ulem}

\newcommand{\xCornell}{Department of Physics, Cornell University, Ithaca, NY, USA}
\newcommand{\xCornellCS}{Department of Computer Science, Cornell University, Ithaca, New York 14853, USA}
\newcommand{\xEwha}{Department of Physics, Ewha Womans University, Seoul, South Korea\\\vspace{0.7em}}
\newcommand{\xGoogle}{Google Research, Mountain View, CA, USA}

\begin{document}
\title{Learning to decode logical circuits}
\author{Yiqing Zhou}
\email{yz2728@cornell.edu}
\affiliation{\xCornell}
\author{Chao Wan}
\affiliation{\xCornellCS}
\author{Yichen Xu}
\affiliation{\xCornell}
\author{Jin Peng Zhou}
\affiliation{\xCornellCS}
\author{Kilian Q. Weinberger}
\affiliation{\xCornellCS}
\author{Eun-Ah Kim}
\email{eun-ah.kim@cornell.edu}
\affiliation{\xCornell}
\affiliation{\xGoogle}
\affiliation{\xEwha}
\begin{abstract}
With the development of quantum hardware bringing the error-corrected quantum circuits to the near future, the lack of an efficient polynomial-time decoding algorithms for logical circuits presents a critical bottleneck. 
While quantum memory decoding has been well-studied, inevitable correlated errors introduced by entangling logical gates prevent the straightforward generalization of quantum memory decoders.
We introduce a data-centric modular decoder framework, Multi-Core Circuit Decoder (MCCD), consisting of decoder modules corresponding to each logical operation supported by the quantum hardware. The MCCD handles 
both single-qubit and entangling gates within a unified framework. We train MCCD using mirror-symmetric random Clifford circuits, demonstrating its ability to effectively learn correlated decoding patterns. Through extensive testing on circuits significantly deeper than those used in training, we show that MCCD maintains high logical accuracy while exhibiting competitive polynomial decoding time across increasing circuit depths and code distances. When compared with conventional decoders like Minimum Weight Perfect Matching (MWPM), Most Likely Error (MLE), and Belief Propagation with Ordered Statistics Post-processing (BP-OSD), MCCD achieves competitive accuracy with substantially better time efficiency, particularly for circuits with entangling gates. Our approach represents a  noise-model agnostic solution to the decoding challenge for deep logical quantum circuits.

\end{abstract}
\maketitle
\para
Recent experimental breakthroughs bring the prospect of logical circuits with quantum error correction
to the near future \cite{Acharya2023Naturea,Bluvstein2024Naturea, Lacroix2024}. 
Quantum error correction codes protect logical information by \textit{encoding} a logical qubit into multiple physical qubits and correcting when the logical information is corrupted. 
The error syndromes from sequential pairs of stabilizer measurements must be processed by a \textit{decoder} to determine when a logical error occurs \cite{Shor1995Phys.Rev.A,Calderbank1996Phys.Rev.A,Gottesman1997,Knill1997Phys.Rev.A}. However, decoding is itself a challenging task for classical computers. For instance, calculating the most likely error is an NP-hard problem \cite{Berlekamp1978IEEETrans.Inf.Theory, Vardy1997IEEETrans.Inf.Theory} for the surface code~\cite{Fowler2012Phys.Rev.A}.
Conventional approaches construct noise-model dependent decoding graphs and leverage the available efficient graph solvers~\cite{Fowler2012Phys.Rev.Lett.,Higgott2022ACMTransactionsonQuantumComputing}. 
However, correlated errors introduce hyperedges
that are beyond the reach of matching-based graph solvers~\cite{Fowler2013,Cain2024Phys.Rev.Lett.} without noise model-driven decomposition
\cite{Delfosse2014Phys.Rev.A,Paler2023Quantum,Sahay2024,Wan2024}. 
Since errors inevitably propagate between logical qubits under entangling logical gates, the challenge of correlated errors becomes a key bottleneck issue for decoding logical circuit operations. 
Hence, an effective and noise model-agnostic decoder for logical circuits approach to decoding correlated errors is critical for creating entangled fault-tolerant states. 

\begin{figure*}[h]
    \centering
    \includegraphics[width=1.9\columnwidth]{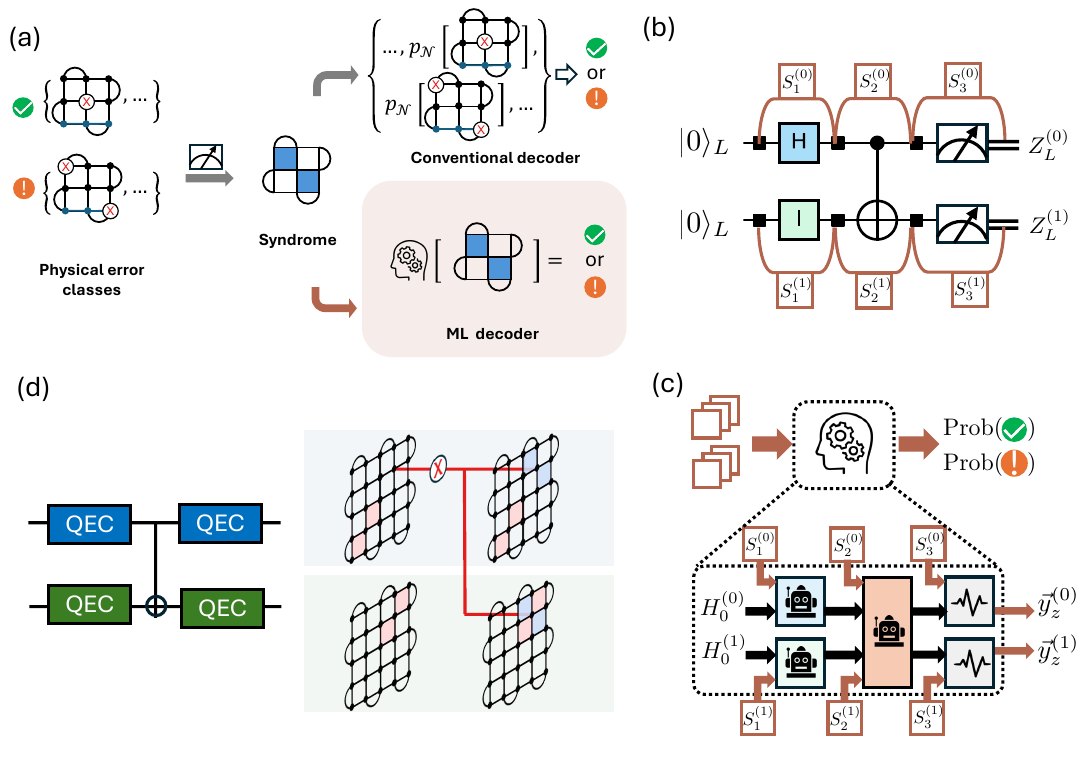}
    \caption{
    (a) The decoding problem statement. Decoding can be viewed as a binary classification problem as demonstrated in a $d=3$ surface code, whose $Z$ logical operator is chosen to be the product of the three blue physical qubits. A given error syndrome (where the violated $Z$ stabilizers are marked in blue squares) may be caused by two classses of physical $X$ errors, whose locations are marked by ``$X$" on the physical qubits. Physical $X$ errors in class ``0" do not cause logical $Z$ error, while errors in class ``1" flip the logical $Z$. Conventional manual decoders aim to reconstruct the physical error from the syndrome, and the binary decision of 0 or 1 is made through the physical error probability $p_\mathcal{N}$ with with a given noise model $\mathcal{N}$. In contrast, the pre-trained ML decoder directly performs a binary classification task using only the syndrome data.
    (b) A sketch of the logical circuit that prepares a logical Bell pair. Both logical qubits are initialized in $|0\rangle_L$ state in logical $Z$ basis. The black squares represents QEC rounds where stabilizers are measured. Syndromes are constructed from neighboring stabilizer measurements.  
    (c) A ML decoder takes sequences of syndromes from different logical qubits and predicts whether logical error happened for each logical qubit. Our ML-based decoder is composed of a decoding module network that parallels the corresponding logical circuits. 
    (d) Entangling logical operations propagats physical error from one logical qubit to the other. As marked by the red lines, a physical $X$ error on the control logical qubit (top) can propagate through a logical CNOT gate to the target logical qubit (bottom). The syndromes of control and target qubit are correlated and both reflect the original physical $X$ error in the control qubit. 
    }
    \label{fig:decoding_problem_statement}
\end{figure*}

\para From a data-centric perspective, the objective of a decoder is to determine whether a given syndrome data stems from a logical error [see Fig.~\ref{fig:decoding_problem_statement}(a)]. The challenge is that many physical error configurations manifest an identical syndrome. 
Conventional approaches heads-on face the under-constrained inverse problem by manually deriving a probabilistic model for the physical error configurations associated with a given syndrome, relying on the noise model. 
Driven by the need for noise-model agnostic and scalable decoders, there have been efforts to develop machine learning (ML) based decoders. ML  decoders frame decoding as a classification problem, aiming directly at the binary prediction of logical error occurrence. By avoiding explicit reconstruction of physical errors, ML decoder can be more scalable than conventional approaches. 
However, most ML-decoder efforts have been limited to decoding a single logical qubit, such as a quantum memory \cite{Terhal2015Rev.Mod.Phys.,Baireuther2018Quantum, Baireuther2019NewJ.Phys., Varbanov2023} or a qubit under single-qubit logical operations \cite{Lacroix2024}. Here we develop a modular Multi-Core Circuit Decoder (MCCD) custom designed from the data-centric perspective [Fig.~\ref{fig:decoding_problem_statement}(b-c)] and demonstrate its efficacy on data from random Clifford circuits simulated with a realistic noise level.
We find MCCD to learn to decode correlate errors from a general circuit successfully, showing favorable linear scaling in wall-time with circuit depth. 

\begin{figure*}[h]
    \centering
    \includegraphics[width=1.9\columnwidth]{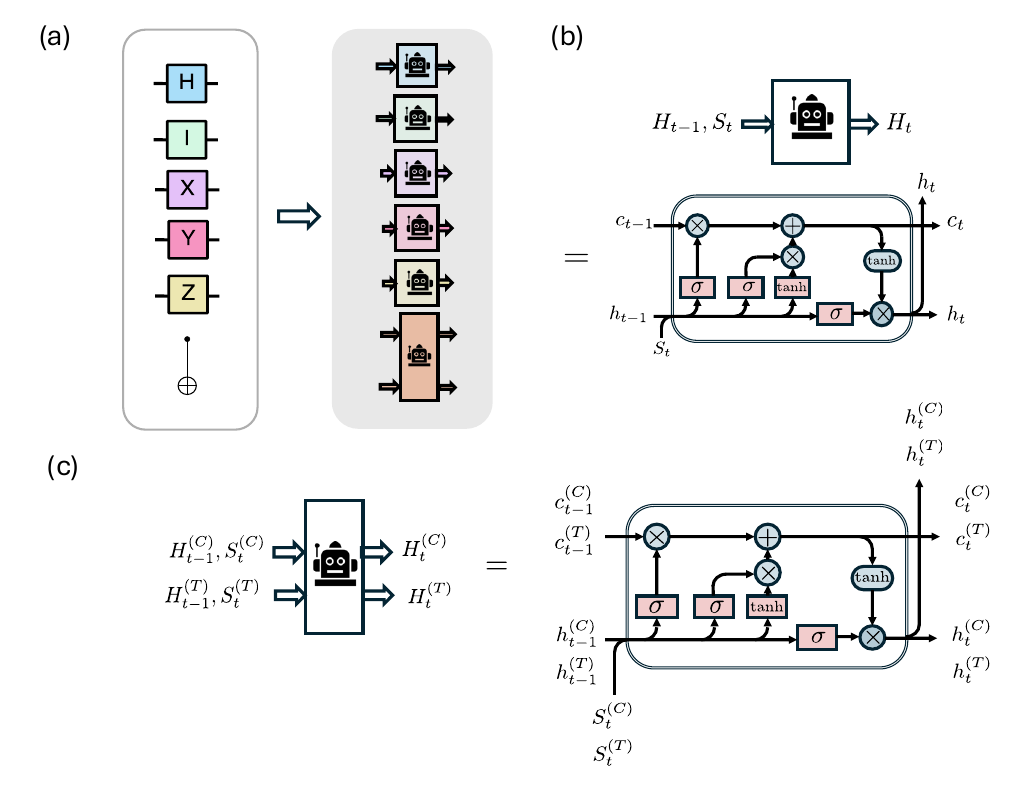}
    \caption{
    ML architecture. (a) For each element in the gate set, we train a separate decoder module. There is a one-to-one mapping between the logical gates (left box) to the ML decoder module (right box). 
    (b) Single-qubit decoder module inner structure based on a LSTM core. The hidden state $H_{t-1}={c_{t-1}, h_{t-1}}$ (we omit superscript $(q)$), has two components $c_{t-1}$ and $h_{t-1}$, which are designed to keep track of long-term and short-term history correspondingly. Together with the new syndrome $S_t$, $H_{t-1}$ follows the internal information flow as show in the box, and output updated hidden state $H_{t}$. The red boxed operations represent linear layer followed by corresponding activation functions (sigmoid $\sigma$ and $\tanh$). The blue circled operations represents element-wise addition ($+$), multiplication ($\times$) and $\tanh$. 
    (c) To handle correlated decoding, we use a module that input the hidden states $H_{t-1}^{(C)}, H_{t-1}^{(T)}$ and new syndromes $S_t^{(C)}, S_t^{(T)}$ from both control $(C)$ and target $(T)$ logical qubits. The hidden states are updated simultaneously. 
    } 
    \label{fig:lstm}
\end{figure*}

\para At a high level, the syndrome data consist of time sequences,  each associated with a logical qubit from quantum error correction rounds. Accumulation of physical noise in each individual code block introduces intra-sequence time correlations. Without gate operations, each time-sequence will be independently time-translation invariant. This structure motivates the use of recurrent structure with a hidden state for each logical qubit to capture the temporal correlation. Indeed, a recurrent neural network~\cite{Rumelhart1986} with a single recurrent {\it cell} was proposed for decoding quantum memory~\cite{Baireuther2018Quantum}. 
Our goal of developing a versatile logical circuit decoder introduces three new challenges: (1) time-translation invariance is broken with gate operations, (2) entangling gates introduce inter-sequence correlations [see Fig.~\ref{fig:decoding_problem_statement}(d)], (3) the decoder should be able to handle any circuits built with a fixed gate set. 
The MCCD tackles all three challenges in one swoop by employing multiple processing cells (PC's), one dedicated processing cell for each logical gate in the gate set [see Fig.~\ref{fig:lstm}(a)]. 
The MCCD handles single-qubit logical gates and entangling logical gates on equal footing by simply feeding syndrome data into the corresponding recurrent cell as the data is collected. The only assumption the MCCD makes is the existence of a native logical gate set for a given quantum hardware, which is a universal assumption for digital quantum computers.

\para 
Specifically, the MCCD uses
one time-dependent hidden state  $H_t^{(q)}$  per logical qubit $q$ to capture its error history and multiple PC's, one for each gate, to build a modularized decoder.  The hidden states serve as a memory to keep track of the error history. The PC's and the readout network are trained for a fixed gate set. 
To decode deep logical circuits, we design each PC based on the recurrent cell of long-short-term-memory \cite{Hochreiter1997NeuralComputation}, which avoids the vanishing gradient problem of standard recurrent neural networks
 when processing long sequences. 
This choice is inspired by the successful memory decoding of up to hundreds of quantum error correction rounds using a long-short-term-memory architecture for a surface code
\cite{Baireuther2018Quantum}.
Hence each hidden state $H_{t-1} = (c_{t-1}, h_{t-1})$ 
  has two components, $c_{t-1}$ and $h_{t-1}$, which track long-term and short-term history, correspondingly. 

\para Before the logical circuit starts to run, 
we initialize the hidden states $H_0^{(q)}$ to all zeros for each logical qubit $q$. 
When each logical gate is applied at time $t$ and a new round of syndrome $S_t$ is extracted, we feed 
the hidden state $H_{t-1}$ and the syndrome data $S_t$ to the gate-specific PC for the cell to output an updated hidden state $H_{t}$.
The PC for an entangling logical gate must capture the physical process of error propagation through the entangling operation. For this, this PC is designed to take hidden states and syndromes from both logical qubits (e.g., the control and target qubits in the case of CNOT gates in Fig.~\ref{fig:lstm}(c)) as 
inputs, and update the two hidden states simultaneously [Fig.~\ref{fig:lstm}(c)]. 
The hidden states provide error history information, while the new syndromes introduce information regarding new physical errors, both necessary for correlated decoding. Finally, a trainable readout network maps the final hidden state to the logical error prediction output $\vec{y}_Z^{(q)}$.
To improve performance, we provide the readout network with the final round syndrome measurement constructed from the data-qubit measurement in addition to the hidden state~\footnote{Only half of the stabilizers can be constructed depending on the measurement basis of the data qubits.}. 

\begin{figure*}[h]
    \centering
    \includegraphics[width=1.8\columnwidth]{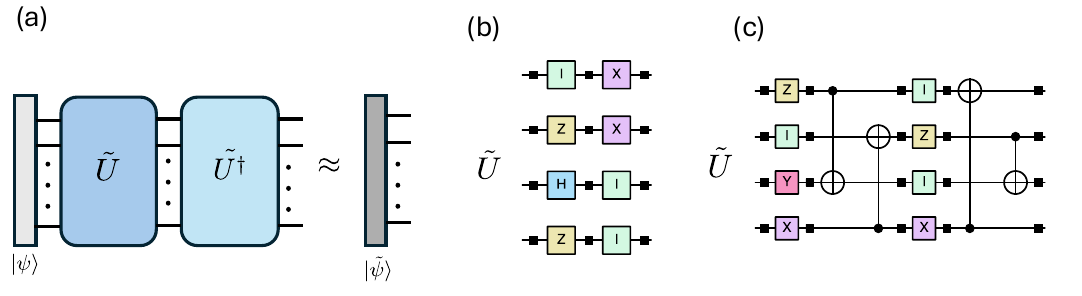}
    \caption{Mirror-symmetric random circuits. (a) A mirror symmetric circuit can be generated for an arbitrary circuit $U$ by concatenating a reversed circuit $U^\dagger$ after the forward evolution $U$. The tilde in $\tilde{U}$ highlights the existence of noise. In the noiseless limit, the mirror symmetry circuit $UU^\dagger$ applied on some initial state $|\psi\rangle$ should keep the state untouched. Due to the noise $\tilde{U}\tilde{U}^\dagger|\psi\rangle = |\tilde{\psi}\rangle \approx |\psi\rangle$. 
    (b) An example of $\tilde{U}$ sampled from \textit{Circuit Type I}. (c)  An example of $\tilde{U}$ sampled from \textit{Circuit Type II}.
}
    \label{fig:mirror_circuit}
\end{figure*}

 \para For MCCD to be effective, the training must be driven by a physically meaningful loss function and diverse training data that avoid overfitting. 
We employ a training scheme inspired by logical randomized benchmarking (RB)~\cite{Combes2017}, as random circuits can prevent MCCD from learning subtle fingerprints of specific circuit used in training. We overcome the difficulty of defining a universal loss function while using random circuit by using mirror-symmetric circuits that evolve a state forward by a random unitary $U$ composed of logical gates in the gate set and then backward by $U^\dagger$, as shown in Fig.~\ref{fig:mirror_circuit}(a). Since noiseless implementation of $U$ and $U^\dagger$ should keep the initial state unchanged 
  , the differences between the final logical measurements and the initial logical preparation offer ground-truth training labels. 
  Hence, we use cross entropy between the model outputs and training labels in such mirror-symmetric circuits as the loss function to drive the training of the MCCD.  To ensure diversity in the training data, we opt for a hardware-efficient implementation of random unitaries. 
 Finally, to ensure that MCCD's training generalizes to new circuits, we train MCCD with data from logical circuit depth $D=2,4,6,8,10$ ($D=4, 8, 12, 16, 20$), and test the model on a wide range of unseen circuits with depth up to $D=18$ ($D=36$) for \textit{Circuit Type I} (\textit{Circuit Type II}), defined below. 

 \para 
 We demonstrate the efficacy and versatility of MCCD by testing MCCD on two types of mirror-symmetric random logical circuits:
 \begin{itemize}
     \item \textit{Circuit Type I: } Single-qubit-only logical circuits on each logical qubit. We sample a sequence of random single qubit gates from the gate set $\{I, X, Y, Z, H\}$, followed by the sequence in the reverse order [See Fig.~\ref{fig:mirror_circuit}(b)].
     \item \textit{Circuit Type II: } Random scrambling logical circuit, consists of interlacing layers of single and two qubit gates. Each single qubit gate layer contains randomly sampled gates from gate set $\{I, X, Y, Z , H\}$. Each two-qubit gate layer consists of CNOT gates entangling random pairs of logical qubits; each logical qubit is act on by only one CNOT. Again, the mirror-symmetry is enforced by running the gate sequences in reverse order [See 
     Fig.~\ref{fig:mirror_circuit}(c)].
 \end{itemize}
We generated the training data using
stim package \cite{Gidney2021Quantum} to simulate the logical circuits based on surface code with the circuit-level noise model motivated by the experimental noise in neutral-atom platforms \cite{Bluvstein2024Naturea, Cain2024Phys.Rev.Lett.}. See the SM Section C for the details of the simulation.
Dividing our study into the above two circuit types achieves two goals. First, it allows us to benchmark against conventional methods on the easier task of decoding circuits ({\it Type I}) without correlated error. Second, it allows  us to
employ a resource-efficient curriculum training strategy\cite{Bengio2009Proc.26thAnnu.Int.Conf.Mach.Learn.}. Specifically, the initial phase of training uses \textit{Type I circuits}, allowing training on data that is cheaper to generate. The resulting modules then serve as initialization for more complicated \textit{Type II circuits}, focusing training effort on capturing correlated error structures induced by entangling logical gates, which are not present in Type I, while optimizing for lower overall data generation cost.
 
\para A meritorious decoder must optimize between two competing needs: accuracy and speed. While a high accuracy is desired, when a decoder's speed lags behind the clock cycle of the quantum computer, it will add a crippling burden. Hence, we benchmark the performance of MCCD from the dual perspectives of the logical accuracy and the wall time, against the performances of three established conventional decoders. Specifically, we consider 
minimum weight perfect matching (MWPM)~\cite{Higgott2022ACMTransactionsonQuantumComputing, Higgott2025Quantum}, most likely error (MLE)~\cite{Cain2024Phys.Rev.Lett.} and belief propagation with ordered statistics post-processing (BPOSD)~\cite{Roffe2020Phys.Rev.Research, Panteleev2021Quantuma}. We compare the decoder performances in decoding logical circuits built using surface code-based logical qubits with two different code distances $d=3,5$ and transversal logical Clifford gates, including $\{I, X, Y, Z, H, \text{CNOT}\}$.  

\begin{figure*}[h]
    \centering
    \includegraphics[width=2.0\columnwidth]{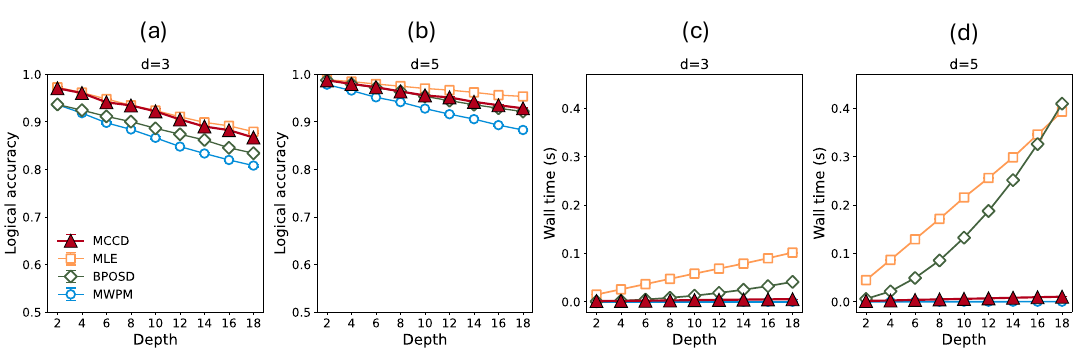}
    \caption{Decoding performance on \textit{Circuit Type I}. (a)-(b) Logical accuracy versus circuit depth using surface code with code distance $d=3$ and $d=5$ for MCCD (red), MLE (yellow), BP-OSD (green), and MWPM (blue) decoders. (c)-(d) Wall time taken to decoding one syndrome trajectory with increasing circuit depth for a $d=3$ and $d=5$ surface code. 
    }
    \label{fig:performance_type_i}
\end{figure*}

\para 
We first train and test MCCD on simulated data from \textit{Circuit Type I} with only single-qubit gates [see Fig.~\ref{fig:mirror_circuit}(a)-(b)]. Although training was limited to the circuits of depths up to $D=10$, neither the logical accuracy [Fig.~\ref{fig:performance_type_i} (a) and (b)] nor the wall time [Fig.~\ref{fig:performance_type_i} (c) and (d)] show any sign of deterioration upon crossing the training depth limit of $D=10$. This indicates that MCCD is indeed capable of decoding circuits much deeper than those used for training. Comparison of MCCD performances between code distances $d=3$ [Fig.~\ref{fig:performance_type_i} (a)] and $d=5$ [Fig.~\ref{fig:performance_type_i} (b)], the improved accuracy of MCCD upon increasing the code distance shows that the noise level we adopted to match the experimental noise level~\cite{Bluvstein2024Naturea} is below the MCCD's decoding threshold. Most remarkably, MCCD's wall time remains much lower than other decoders upon increasing the code distance or the circuit depth [Fig.~\ref{fig:performance_type_i} (c) and (d)].  

\para
When comparing the logical accuracies of the reference decoders for the two code distances $d=3$ and $d=5$, clearly the noise level of the data is below the error threshold for all the decoders under consideration. Since MLE considers all possible error patterns, it achieves higher accuracy, especially at deeper circuits [Fig.~\ref{fig:performance_type_i} (a) and (b)]. However, this accuracy is achieved at the cost of punishing exponentially increasing wall time[Fig.~\ref{fig:performance_type_i} (c) and (d)].
On the other hand, while MWPM runs at high speed that is comparable to the speed of MCCD [Fig.~\ref{fig:performance_type_i} (c) and (d)], it shows significantly lower accuracy compared to MCCD [Fig.~\ref{fig:performance_type_i} (a) and (b)].
BP-OSD offers better accuracy than MWPM, but its wall time sharply increases with the circuit depth for code distance $d=5$. 
In comparison, MCCD shows competitive accuracy with a linear-in-depth wall time scaling. 

\begin{figure*}[h]
    \centering
    \includegraphics[width=2.0\columnwidth]{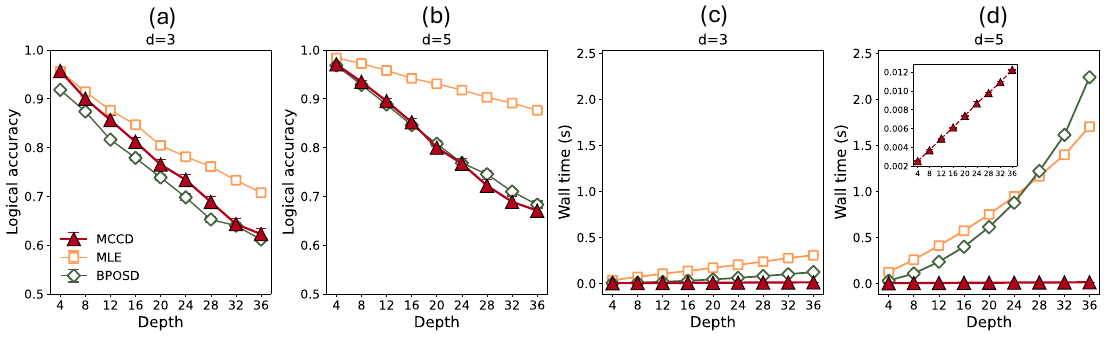}
    \caption{Decoding performance on \textit{Circuit Type II}. (a)-(b) Logical accuracy versus circuit depth using surface code with code distance $d=3$ and $d=5$ for MCCD (red), MLE (yellow) and BP-OSD (green) decoders. (c)-(d) Wall time taken to decoding one syndrome trajectory with increasing circuit depth for a $d=3$ and $d=5$ surface code. The inset of panel (d) shows a zoomed in view of computing time for ML decoder. 
    }
    \label{fig:performance_type_ii}
\end{figure*}

\para 
Building on the above encouraging results on the data from \textit{Type I} circuits, we follow a curriculum learning approach~\cite{Bengio2009Proc.26thAnnu.Int.Conf.Mach.Learn.}, to train an extra module for the logical CNOT gate while keeping single-qubit gate decoder modules and readout network fixed. 
To generate diverse training data that include two-qubit logical gates, we sample mirror-symmetric logical circuits sampled from \textit{Circuit Type II} [Fig.~\ref{fig:mirror_circuit}(a) and (c)]. 
As shown through the improvements in the logical accuracy upon increasing the code distance in Fig.~\ref{fig:performance_type_ii}(a)-(b), the MCCD's correlated decoding threshold is above realistic noise level of our simulated data. Most importantly, the MCCD's wall time remains practically linear as a function of circuit depth, even with the correlated errors due to the entangling gates.  

\para 
By contrast, the MWPM which was comparable to our MCCD in its speed for \textit{Circuit Type I} is incapable of handling correlated errors in the \textit{Circuit Type II} data. Hence, we limit the comparison in Fig.~\ref{fig:performance_type_ii} to MLE~\cite{Cain2024Phys.Rev.Lett.} and BP-OSD~\cite{Roffe2020Phys.Rev.Research,Panteleev2021Quantuma}. Unsurprisingly, MLE shows higher accuracy. However, once again, the exponential cost of MLE~\cite{Cain2024Phys.Rev.Lett.} is prohibitively high.  While BP-OSD is faster than MLE at low circuit depth, its wall time rapidly increases with circuit depth at code distance $d=5$. Hence MCCD with linear complexity $O(D)$ for the logical circuit of depth $D$ clearly outperforms the conventional decoders in running time especially in the large depth region. 

\para In summary, we introduced pre-trainable and modularized MCCD and its efficient training scheme as machine-learning based decoder 
that successfully addresses the challenge of decoding logical quantum circuits with correlated errors. Utilizing a data-centric machine learning approach with dedicated processing cells for each gate type, our decoder effectively captures error propagation through both single-qubit and entangling logical operations. The MCCD's performance demonstrates three key advantages: (1) noise-model agnosticism that allows application to various quantum hardware platforms without detailed knowledge of underlying noise characteristics, (2) high decoding accuracy competitive with conventional decoders, and (3) remarkable practical run time efficiency and promising complexity scaling with circuit depth. This linear scaling behavior with a small prefactor makes MCCD particularly suitable for real-time decoding in practical quantum computing environments with deep logical circuits. Furthermore, our modular design enables straightforward extension to new gate sets or improved physical implementations of existing gates as quantum hardware evolves. The combination of accuracy, speed, and adaptability positions MCCD as a promising approach for facilitating the path toward fault-tolerant quantum computation.

\para Interesting directions of future work include the extension of MCCD to include mid-circuit measurement and feedforward operations \cite{zhou2024} and to larger code distances. While we
 focused on the transversal implementation of logical Clifford gate~\cite{Gottesman1997,Bombin2006Phys.Rev.Lett.,Gottesman2014} in this work, inherent flexibility of MCCD would allow for the extension of the set of logical operations. Such extension will open doors to teleportation-based implementation of non-Clifford logical gates \cite{Knill2004,Bravyi2005Phys.Rev.A,Bravyi2012Phys.Rev.A} and lattice surgery implementation of entangling gates \cite{Horsman2012NewJ.Phys.,Litinski2018Quantum}. 
 Scaling up the code distance will be necessary for real-time decoding of fault-tolerant logical circuits. Strategies developed by the machine learning community for increasing the syntax window for large language models could be productive as the syndrome sequence is a time sequence. Furthermore, strategic adoption of progress in language models can enable expansion of the syndrome time window of MCCD to capture longer time-scale temporal correlations.   

\paragraph{Note - } Toward the conclusion of this work, we noticed a recent work (Ref. ~\cite{Lacroix2024}) used a similar approach to train a neural network decoder for the color code, demonstrating its effectiveness on single-qubit Clifford logical circuits. 

\paragraph{Code availability - } The code supporting the findings of this study is available at \href{https://github.com/KimGroup/MCCD.git}{https://github.com/KimGroup/MCCD.git}. 
\paragraph{Acknowledgements - }  Y.Z. thanks Madelyn Cain for sharing the simulation code used in Ref.~\cite{Cain2024Phys.Rev.Lett.}. H.K., C.W., Y.X., J.Z., K.Q.W. , and E-A.K. acknowledge support from the NSF through OAC-2118310. Y.Z. acknowledges support from NSF Materials Research Science and Engineering Center (MRSEC) through DMR-1719875 and from Platform for the Accelerated Realization, Analysis, and Discovery of Interface Materials (PARADIM), supported by the NSF under Cooperative Agreement No.\ DMR-2039380. The computation was carried out on the Cornell G2 cluster established in part with the support from the Gordon and Betty Moore Foundation’s EPiQS Initiative, Grant GBMF10436 to  E-AK.

\bibliography{bibliography}

\newpage

\end{document}


\title{Supplementary Materials for ``Learning to decode logical circuits"}
\author{Yiqing Zhou}
\affiliation{\xCornell}
\author{Chao Wan}
\affiliation{\xCornellCS}
\author{Yichen Xu}
\affiliation{\xCornell}
\author{Jin Peng Zhou}
\affiliation{\xCornellCS}
\author{Kilian Q. Weinberger}
\affiliation{\xCornellCS}
\author{Eun-Ah Kim}
\affiliation{\xCornell}
\affiliation{\xGoogle}
\affiliation{\xEwha}

\date{\today}

\maketitle
\tableofcontents
\appendix
\section{Architecture}
Our modular ML-based decoder consists of three types of modules: (1) decoder modules for single qubit logical operations, (2) decoder module for the entangling logical operation, (3) a readout module in logical $Z$ basis. As mentioned in the main text, we consider measurements in $Z$ logical basis to demonstrate the efficacy of our modular ML-based decoder. However, one can handle the $X$ basis logical measurements by either replacing the $X$ logical measurement with an $\bar{H}$ followed by logical $Z$ measurement using our pre-trained decoder modules, or train an extra readout network for $X$ logical measurements. 
In the following, we discuss the architecture of the three types of modules. 

\paragraph{Decoder modules for single qubit logical operations}
To effectively learn how to decode logical circuits, we observe that different logical operations transform error history differently and potentially introduces different type of physical errors depending on the physical implementation of the logical gate. Therefore, we need to train a different decoder module for different logical operations. We utilize different recurrent modules to process the hidden state of a surface code at a certain time point according to the gate operation at this time point. Therefore, the number of single qubit LSTM modules needed is equal to the number of different single qubit gates in the circuit. Each single qubit LSTM module takes the hidden state from previous time step and outputs the hidden state for the next time step. 

We use LSTM cell with $2$ layers in the single qubit decoder modules. Mathematically, each layer in the LSTM module processes the hidden state at time step $t$ in the following way: 
\begin{equation}
    \begin{split}
        i_t &= \sigma (W_{ii}x_t + b_{ii} + W_{hi}h_{i-1}+b_{hi}) \\
        f_t &= \sigma (W_{if}x_t + b_{if} + W_{hf}h_{i-1}+b_{hf}) \\
        g_t &= \tanh (W_{ig}x_t + b_{ig} + W_{hg}h_{i-1}+b_{hg}) \\
        o_t &= \sigma (W_{io}x_t + b_{io} + W_{h}h_{i-1}+b_{ho}) \\
        c_t &= f_t \odot c_{t-1} + i_t \odot g_t \\
        h_t &= o_t \odot \tanh(c_t)
    \end{split}
\end{equation}

In the context of our logical circuit decoder, $x_t$ is the new syndrome obtained at the current time step; $c_{t-1}$ and $h_{t-1}$ together, $H_{t-1} = (c_{t-1}, h_{t-1})$, is the hidden state of the logical qubit in the previous time step $t-1$. The output of the module is an updated hidden state $H_t = (c_t, h_t)$.

Unlike regular RNNs, LSTMs introduce a memory cell ($c_t$) that allows information to be stored over long sequences. The memory cell is controlled by three gates: forget gate, input gate and output gate which decides what information should be removed from the memory cell, added to the memory, and passed to the next step respectively. 

The input to the model contains three parts: $h_t$, the hidden state (short term memory) at time step $t$; $x_t$, the input at time step $t$, corresponding to the syndrome measurement in our decoding problem; and $c_t$, the cell state (long-term memory) for the RNN core. The weight ($W$) and bias ($b$) matrices for each gate are multiplied to $x_t$ and $h_t$ to control how the next step is updated. The output from the LSTM module contains two parts: an updated hidden state for the next time step $h_{t+1}$ and an updated cell state $c_{t+1}$. We only take the hidden state from the output of the previous module and feed it as the hidden state for the next module. Since we use different single qubit LSTM modules for different time steps depending on the single qubit gate, the recurrent structure is different from the traditional recurrent LSTM core. 

\paragraph{Two qubit LSTM module}
To model the entanglement through a two qubit gate between two previously non-correlated surface codes, we introduce the two qubit LSTM module from the onset of entanglement. 

The input of the two qubit LSTM module is the concatenated hidden state from the two separate codes. Since the physical gate operation on the qubits is not symmetrical, we need to keep the order or concatenation consistent with the order of the gate operation. For the sake of simplicity, we always put the hidden state with the control qubit at the first place. Since we are using the same architecture for the two qubit operation, the two qubit hidden state transformed similarly as the single qubit hidden state. The difference is the dimension of the hidden state is twice as large. From the output of the two qubit LSTM module, we split the final hidden state into two in the hidden dimension, and take the first one as the updated hidden state for the surface code containing the control qubit, and the other half as the updated hidden state for the surface code with the target qubit.

\paragraph{Auxiliary Readout}
The auxiliary readout module aims to output the probability of error for a logical qubit. It consists of two linear layers connected by ReLU activation between them. Since we are dealing with multiple logical qubits, we process the single logical qubit hidden state first and concatenate after the readout module. We use the same linear layers for different types of logical qubits but process them accordingly before the readout. 

\paragraph{Main Readout}
Similar to the auxiliary readout module, the main readout also serves as a classification module that outputs the error probability of the current logical qubit. The difference lies in the input. The syndrome measurement from the final round, which only contains half of the stabilizers constructed from the final data qubit measurements, is concatenated to the hidden state representation before fed into the main readout. 

\paragraph{Loss function}
The total loss of our decoder is a weighted sum of the cross entropy loss of the main and the auxiliary readout module. We choose to weight the auxiliary loss by a factor of 0.5~\cite{varbanovNeuralNetworkDecoder2023}. 

\section{MCCD Training}
\paragraph{Training pipeline}
There are two intertwined tasks in the error correction for correlated surface code problem: gate-dependent error correction for single qubit and correlated error correction for two qubits for the CNOT gate. We observe that separating the two tasks and training the model sequentially on the two types of modules make the learning process easier and lead to promising result. 
Therefore, we utilize a two-stage training strategy by first training the single-qubit related modules in \textit{Circuit Type I} and then training the two-qubit LSTM core only in \textit{Circuit Type II}. In the first stage of training, we optimize parameters in all the single-qubit related modules. These include all the recurrent LSTM modules for single-qubit, the main readout, the auxiliary readout. After the single-qubit modules are fully trained, we introduce the two-qubit module into the architecture and freeze everything else except the two-qubit processing cell which takes the hidden states of the two entangled logical qubits.  

\paragraph{Training Hyperparameters} The detailed hyperparameters used in the network for training a MCCD for code distance $d=3$ and $d=5$ are listed in Table~\ref{tab:decoder_hyperparams_d3}, and Table~\ref{tab:decoder_hyperparams_d5} correspondingly. The hyperparameters used in training are listed in Table ~\ref{tab:training_hyperparams_summary}. Please also see the published code~\footnote{Code available at \href{https://github.com/KimGroup/MCCD.git}{https://github.com/KimGroup/MCCD.git}}. 

\begin{table}[htbp]
\caption{\label{tab:decoder_hyperparams_d3}
Model hyperparameters used for the staged-trained \texttt{CircuitLSTMDecoderw2Q} for code distance $d=3$.
}
\begin{ruledtabular}
\begin{tabular}{llp{7.5cm}}
\textrm{Hyperparameter} &
\textrm{Value} &
\textrm{Description} \\
\colrule
Input size ($\texttt{input\_size}$) & 8 & Dimensionality of input features per qubit per time step. \\
Hidden size (1Q, $\texttt{hidden\_size}$) & 64 & Hidden state size for all single-qubit LSTM modules. \\
Hidden size (2Q) & 128 & Hidden state size for the joint two-qubit LSTM module. \\
LSTM layers ($\texttt{num\_layers}$) & 2 & Number of stacked layers in each LSTM unit. \\
Two-qubit input size & 16 & Input dimensionality to the two-qubit LSTM (concatenation of two 8-d vectors). \\
Auxiliary input length ($\texttt{fx\_len}$) & 4 & Length of final-round syndrome vector per qubit; concatenated in main readout. \\
Activation function & ReLU & Applied to LSTM output before readout layers. \\
Main readout & Linear(68 $\rightarrow$ 68 $\rightarrow$ 2) & Feedforward network over LSTM output and auxiliary input. \\
Auxiliary readout & Linear(64 $\rightarrow$ 64 $\rightarrow$ 2) & Feedforward network over LSTM output only. \\
Output classes & 2 & Number of classes per logical qubit (binary classification). \\
Gate-dependent LSTM modules & $\{$I, X, Y, Z, H$\}$ & Each gate is modeled by a dedicated single-qubit LSTM module. \\
Two-qubit gate module & \texttt{TwoQubitLSTM} & Shared LSTM applied to fused features and hidden states of qubit pairs. \\
\end{tabular}
\end{ruledtabular}
\end{table}

\begin{table}[htbp]
\caption{\label{tab:decoder_hyperparams_d5}
Model hyperparameters used for the scaled-up \texttt{CircuitLSTMDecoderw2Q} at code distance $d=5$.
}
\begin{ruledtabular}
\begin{tabular}{llp{7.5cm}}
\textrm{Hyperparameter} &
\textrm{Value} &
\textrm{Description} \\
\colrule
Input size ($\texttt{input\_size}$) & 24 & Dimensionality of input features per qubit per time step. \\
Hidden size (1Q, $\texttt{hidden\_size}$) & 192 & Hidden state size for all single-qubit LSTM modules. \\
Hidden size (2Q) & 384 & Hidden state size for the joint two-qubit LSTM module. \\
LSTM layers ($\texttt{num\_layers}$) & 2 & Number of stacked layers in each LSTM unit. \\
Two-qubit input size & 48 & Input dimensionality to the two-qubit LSTM (concatenation of two 24-d vectors). \\
Auxiliary input length ($\texttt{fx\_len}$) & 12 & Length of final-round syndrome vector per qubit; concatenated in main readout. \\
Activation function & ReLU & Applied to LSTM output before readout layers. \\
Main readout & Linear(204 $\rightarrow$ 204 $\rightarrow$ 2) & Feedforward network over LSTM output and auxiliary input. \\
Auxiliary readout & Linear(192 $\rightarrow$ 192 $\rightarrow$ 2) & Feedforward network over LSTM output only. \\
Output classes & 2 & Number of classes per logical qubit (binary classification). \\
Gate-dependent LSTM modules & $\{$I, X, Y, Z, H$\}$ & Each gate is modeled by a dedicated single-qubit LSTM module. \\
Two-qubit gate module & \texttt{TwoQubitLSTM} & Shared LSTM applied to fused features and hidden states of qubit pairs. \\
\end{tabular}
\end{ruledtabular}
\end{table}

\begin{table}[htbp]
\caption{\label{tab:training_hyperparams_summary}
Training hyperparameters.
}
\begin{ruledtabular}
\begin{tabular}{llp{8.5cm}}
\textrm{Hyperparameter} &
\textrm{Value} &
\textrm{Description} \\
\colrule
Optimizer & Adam & Optimization algorithm used for training. \\
Learning rate & 0.001 & Initial learning rate for the Adam optimizer. \\
Loss function & Main CE + 0.5 $\times$ Aux CE & Weighted sum of main and auxiliary cross-entropy losses. \\
Batch size & 1024 & Number of training samples per batch. \\
Training data size & 500{,}000 & Total number of training batches. With batch size 1024, this corresponds to $5 \times 10^8$ syndrome trajectories. \\
\end{tabular}
\end{ruledtabular}
\end{table}

\section{Data generation}
\subsection{The surface code}
\begin{figure*}[h]
    \centering
    \includegraphics[width=0.4\columnwidth]{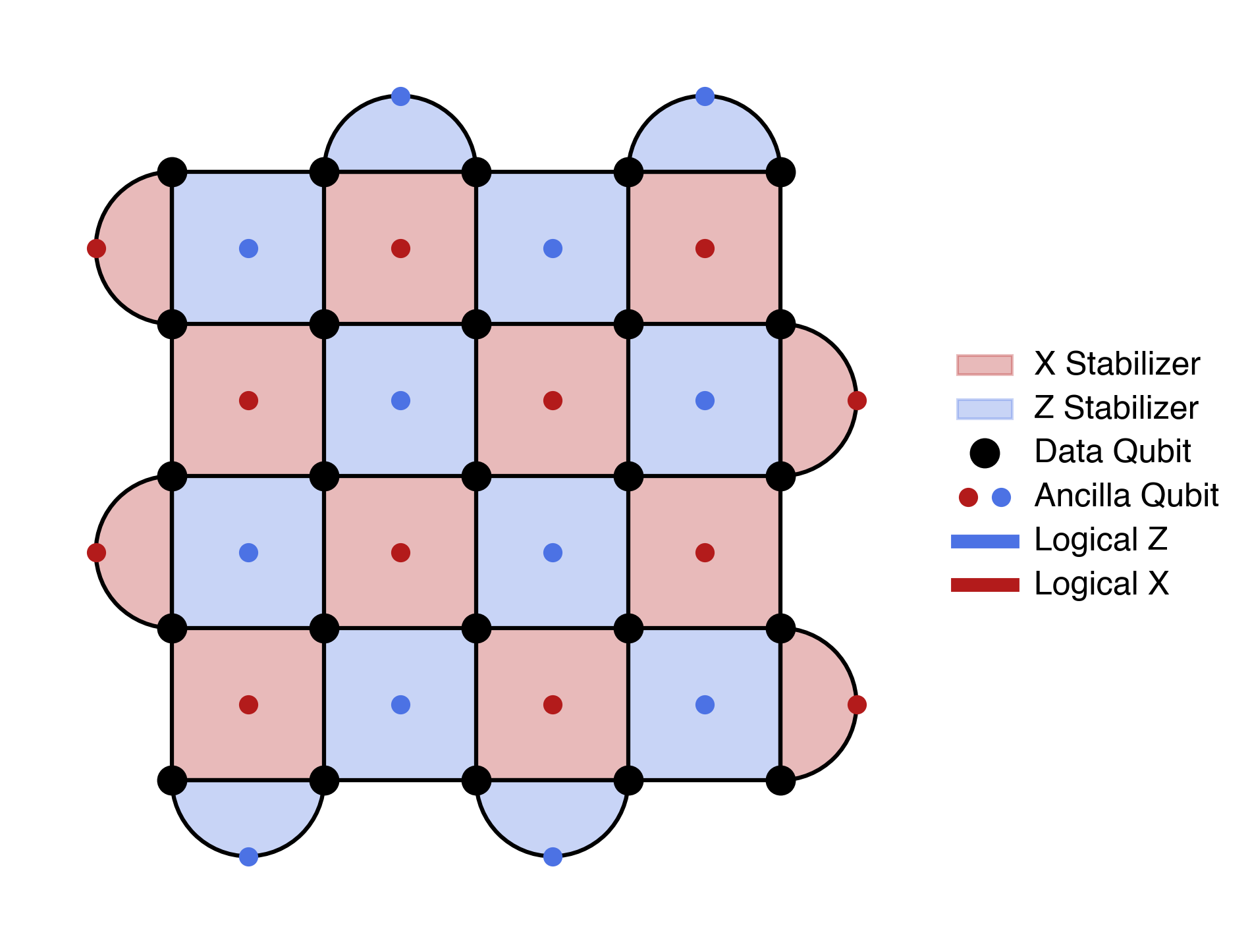}
    \caption{Sketch of a rotated surface code with code distance $d=5$. }
\label{fig:rotated_curface_code}
\end{figure*}

We consider a rotated surface code. An illustration of the code layout is shown in SFig.~\ref{fig:rotated_curface_code}. 
\subsubsection{Logical gate implementations} 
We focus on Clifford logical gates with transversal implementations in this work. For clarity, we denote logical operations with an overbar, $\bar{U}$ and physical operations without an overbar $U$. 

For single qubit logical operations, we consider $\bar{U} \in \{\bar{I}, \bar{X}, \bar{Y}, \bar{Z}, \bar{H}\}$. Each single qubit logical gate $\bar{U}$ corresponds to a depth-$1$ physical circuit where each physical qubit is acted on by physical gate $U$. The Hadamard gate $\bar{H}$ swaps the $\bar{Z}$ and $\bar{X}$ basis, which requires rotating the surface code after applying $H$ on all physical qubits. This rotation can be done ``virtually" by keeping track of the rotation on a classical software, so that it can be done effectively noiselessly on the quantum hardware.

For two-qubit gate, we consider the CNOT gate also implemented as a transversal operation. Physically, the two surface codes are moved to adjacency and physical CNOT gates are applied pair-wise on the physical qubits between the control and target logical qubits. 

\subsubsection{Stabilizer measurement physical circuits}
\begin{figure*}[h]
    \centering
    \includegraphics[width=\columnwidth]{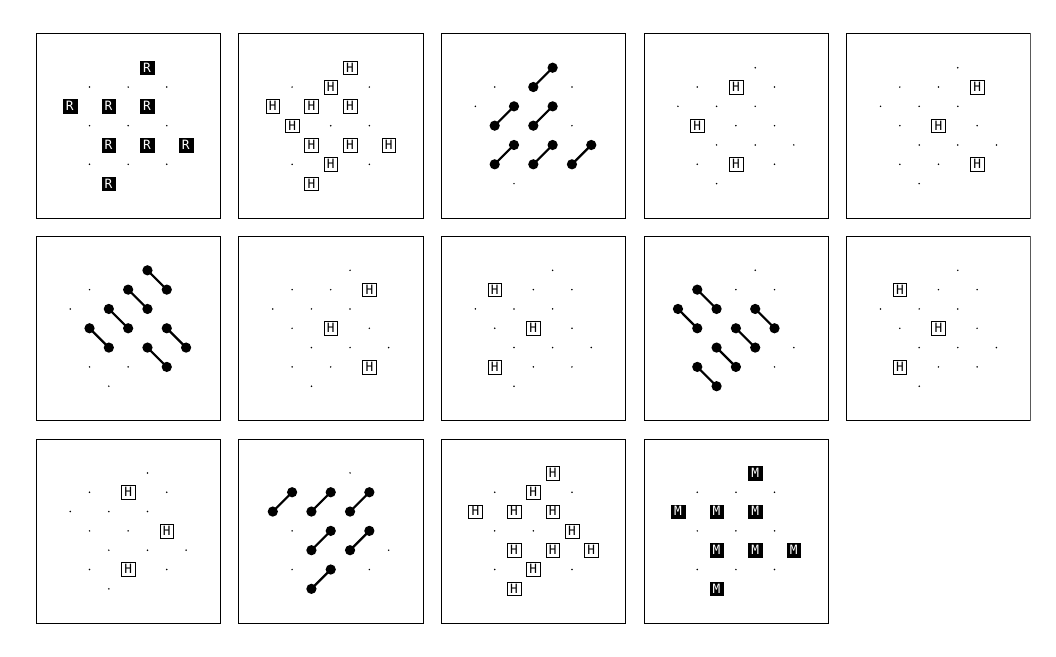}
    \caption{ Visualization of stabilizer measurement circuit. The $R$, $H$, $M$ represents single-qubit reset in $Z$ basis, Hadamard gate, and measurement in $Z$ basis respectively. The two-qubit gates, indicated by dumbbell shaped icons,  are CZ gates with data qubit being the control qubit and ancilla qubits being the target qubit.
    }
    \label{fig:stabilizer_measurement_circuit}
\end{figure*}

The stabilizer measurement circuit couples data qubits with the ancilla qubits. Measuring the ancilla qubits gives us the stabilizer measurement records, which are further processed to obtain syndromes and input into our ML decoder model. This circuit, as illustrated in SFig.~\ref{fig:stabilizer_measurement_circuit}, is applied after each logical gate operation.

\subsection{Simulation noise model}
\label{sm_sec:noise_model}
 For the numerical studies presented in the main text, we use the \texttt{stim} package for simulation. We consider a circuit-level noise model motivated by the current experimental capability of neutral atom array based quantum computers. Specifically, we use a circuit level noise model that includes the following physical noises:
\begin{itemize}
    \item Each two qubit physical gate is followed by a two qubit Pauli noise channel with probability\\
    $[0.0005, 0.00175, 0.000625,$ $ 0.0005, 0, 0, 0, 
    0.00175, 0, 0, 0, 0.000625, 0, 0, 0.00125]$
    \item Each single qubit physical gate is followed by a single qubit depolarizing model with probability \\$[0.0001, 0.0001, 0.0001] $
    \item On a physical level, the atoms are moved to achieve flexible connectivity between different physical qubits. This comes at the cost of having idling error due to the extra time taken during the physical qubit movement, which is captured as a Pauli noise channel with probability $[4\times 10^{-7}, 4\times 10^{-7}, 1.6\times 10^{-6}] $. This error channel is applied when physical qubit movement happens.
    \item Resetting a physical qubit has a bit flip error probability of $p=0.002$.
    \item Measuring a physical qubit has a bit flip error probability of $p=0.002$.
\end{itemize}
\section{Other decoders}
We benchmark MCCD with three different popular decoders, namely, MLE, BPOSD, and MWPM. Here we specify the numerical details of these methods. 
\paragraph{MWPM} We use the open sourced library \texttt{PyMatching}~\cite{higgott2023sparse} with the noise model used for data generation as detailed in SM Sec. ~\ref{sm_sec:noise_model}. 
\paragraph{BPOSD} We use open source library \texttt{stimbposd}~\cite{higgott_stimbposd}. We use the exact noise model used for data generation and maximal belief propagation iterations to be $20$. 
\paragraph{MLE} We use the algorithm developed and implemented as in Ref.~\cite{cainCorrelatedDecodingLogical2024}.  

\clearpage
\input{supp.bbl}

%% file: supp.bbl
%